\documentclass[aps,showpacs,twocolumn]{revtex4}
\usepackage{epsfig}

\begin{document}

\title{The effect of hidden color channels on nucleon-nucleon interaction}

\author{Hongxia Huang$^1$, Pu Xu$^2$, Jialun Ping$^1$ and Fan Wang$^3$}

\affiliation{$^1$Department of Physics, Nanjing Normal University, Nanjing
210097, P.R. China \\
$^2$Department of Applied Physics, Nanjing University of Science and Technology,
 Nanjing 210094, P.R. China \\
$^3$Department of Physics, Nanjing University, Nanjing 210093, P.R. China}

\begin{abstract}
In the framework of constituent quark model, the effect of hidden
color channels on the nucleon-nucleon ($NN$) interaction is studied.
By adjusting the color confinement strength between the hidden color
channels and color singlet channels and/or between the hidden color
channels and hidden color channels, the experimental data of $S$ to
$I$ partial-wave phase shifts of $NN$ scattering can be fitted well.
The results show that the hidden color channel coupling might be
important in producing the intermediate-range attraction of $NN$
interaction. The deuteron properties and dibaryon candidates have
also been studied with this model .
\end{abstract}


\maketitle

\setcounter{totalnumber}{5}

\section{\label{sec:introduction}Introduction}

The study of nucleon-nucleon ($NN$) interaction has lasted over
seventy years. The quantitative description of $NN$ interaction has
been achieved in the one-boson-exchange (OBE) models, the chiral
perturbation theory(ChPT) and quark models. The $\chi^2/$dof$\sim 1$
for more than 2000 data has been obtained in meson exchange
model~\cite{OBE,CPT} and $<$2 in quark model~\cite{QM}.

In the OBE model~\cite{OBE}, the long-range part of the $NN$
interaction is attributed to one-pion-exchange. The short-range part
is described by $\rho,\omega$-meson exchange or phenomenological
repulsive core. While the $\sigma$-meson exchange is responsible for
the intermediate-range attraction. Phenomenological form factors are
needed to achieve the quantitative description of the $NN$
interaction data. In the chiral perturbation theory~\cite{CPT}, the
multi-$\pi$'s are exchanged between two nucleons. The short range
part related to the nucleon internal structure is modeled by the
contact terms with phenomenological low energy constants. The theory
can give a quantitative description of the low-energy $NN$
scattering below the $\pi$ production threshold. It is hard to
extend this model to higher energy, the very interesting resonance
region of $NN$ scattering.

With the advent of quantum chromodynamics (QCD), it is expected to
describe the $NN$ interaction from the fundamental degree of freedom
of QCD, quark and gluon. Recently, lattice QCD calculation has
achieved a qualitative description of $NN$ interaction~\cite{lqcd}.
However it is still far from the quantitative description. The
QCD-inspired quark models are useful in describing the $NN$
interaction with the fundamental quark-gluon degree of freedom. The
most popular and successful one is the constituent quark model.
Where the non-perturbative (color confinement and spontaneous chiral
symmetry breaking) and perturbative properties of QCD are
incorporated into the model by introducing the phenomenological
confinement potential, Goldstone-boson exchange and effective one
gluon exchange between the massive constituent
quarks~\cite{manohar}. Almost in all realistic quark models aimed to
describe the $NN$ interaction, the short-range repulsion of $NN$
interaction is described by one-gluon-exchange and quark
anti-symmetrization. The long-range part is described by $\pi$-meson
exchange which is the same as the OBE and chiral perturbation theory
approaches. To describe the intermediate-range part, the
$\sigma$-meson exchange is employed in most quark model approaches.
The only one exception is the quark delocalization color screening
model (QDCSM). Where the quark delocalization and color screening
effect between interacting quarks within different quark clusters
are employed~\cite{QDCSM} to describe the intermediate range
attraction which is similar to the molecular covalent bond. To
develop such a molecular covalent bond like model is because of the
outstanding fact that the molecular force and nuclear force are
similar except the energy and length scale
difference~\cite{Anderson}. Also because of the existence of
$\sigma$ meson is not sure for long. Recently BES collaboration
reported the observation of $\sigma$-meson, which is appeared as
$\pi\pi$ $S$-wave resonance~\cite{BES}. However, the calculation of
the correlated $\pi\pi$ exchange between two nucleons can not obtain
enough attraction~\cite{sigma} as the phenomenological $\sigma$
meson exchange did. The recent QDCSM calculation, on the other hand,
showed that the quark delocalization and color screening mechanism
is quantitatively equivalent to the phenomenological $\sigma$ meson
exchange in describing the $NN$ intermediate range
attraction~\cite{Chen_NN}. In ChPT there is also no $\sigma$ meson
exchange. In addition, by introducing the multi-body color
confinement interaction~\cite{Oka}, or by incorporating the hidden
color channels in the calculation~\cite{Maltman}, the
intermediate-range attraction can also be obtained to some extent.
Therefore the mechanism of the $NN$ intermediate-range attraction is
still an open question.

In this work, an alternative approach for $NN$ interaction is
studied. The hidden color channels ignored in the prevailing quark
model calculations of $NN$ interaction is included. Accordingly the
confinement potential between different channels is modified as
follows: the ordinary confinement is used for the quark-pairs within
the same nucleon and the color singlet channels whereas a
multiplying factor is introduced for the confinement potential
between the quark pairs if a hidden color channel is involved. The
aim is to test if the color screening phenomenology used in QDCSM is
an effective description of the hidden color channel coupling. The
details of this model approach will be explained in next section.
The $NN$ scattering phase shifts obtained in this approach are
confronted with experimental data and compared with ChQM and QDCSM
approaches. The equivalence of these three quark models in
describing the $NN$ scattering data has been confirmed. The deuteron
properties and dibaryon candidates are also studied with this model.

The structure of this paper is as follows. A brief introduction of
three quark models used is given in
section II. Section III devotes to the numerical results and
discussions. The summary is shown in the last section.

\section{Three quark models}
\subsection{Chiral quark  model}

The Salamanca version of ChQM is chosen as the representative of the
chiral quark models. It has been successfully applied to hadron
spectroscopy and $NN$ interaction. The model details can be found in
Ref.\cite{ChQM1}. Only the Hamiltonian and parameters are given
here. The ChQM Hamiltonian in the nucleon-nucleon sector is
\begin{widetext}
\begin{eqnarray}
H &=& \sum_{i=1}^6 \left(m_i+\frac{p_i^2}{2m_i}\right) -T_c
+\sum_{i<j} \left[
V^{G}(r_{ij})+V^{\pi}(r_{ij})+V^{\sigma}(r_{ij})+V^{C}(r_{ij})
\right],
 \nonumber \\
V^{G}(r_{ij})&=& \frac{1}{4}\alpha_s {\mathbf \lambda}_i \cdot
{\mathbf \lambda}_j
\left[\frac{1}{r_{ij}}-\frac{\pi}{m_q^2}\left(1+\frac{2}{3}
{\mathbf \sigma}_i\cdot {\mathbf\sigma}_j \right)
\delta(r_{ij})-\frac{3}{4m_q^2r^3_{ij}}S_{ij}\right]+V^{G,LS}_{ij},
\nonumber \\
V^{G,LS}_{ij} & = & -\frac{\alpha_s}{4}{\mathbf \lambda}_i
\cdot{\mathbf \lambda}_j
\frac{1}{8m_q^2}\frac{3}{r_{ij}^3}[{\mathbf r}_{ij} \times
({\mathbf p}_i-{\mathbf p}_j)] \cdot({\mathbf \sigma}_i+{\mathbf
\sigma}_j),
\nonumber \\
V^{\pi}(r_{ij})&=& \frac{1}{3}\alpha_{ch}
\frac{\Lambda^2}{\Lambda^2-m_{\pi}^2}m_\pi \left\{ \left[ Y(m_\pi
r_{ij})- \frac{\Lambda^3}{m_{\pi}^3}Y(\Lambda r_{ij}) \right]
{\mathbf \sigma}_i \cdot{\mathbf \sigma}_j \right.\nonumber \\
&& \left. +\left[ H(m_\pi r_{ij})-\frac{\Lambda^3}{m_\pi^3}
H(\Lambda r_{ij})\right] S_{ij} \right\} {\mathbf \tau}_i \cdot {\mathbf \tau}_j,  \\
V^{\sigma}(r_{ij})&=& -\alpha_{ch} \frac{4m_u^2}{m_\pi^2}
\frac{\Lambda^2}{\Lambda^2-m_{\sigma}^2}m_\sigma \left[ Y(m_\sigma
r_{ij})-\frac{\Lambda}{m_\sigma}Y(\Lambda r_{ij})
\right]+V^{\sigma,LS}_{ij}, ~~~~
 \alpha_{ch}= \frac{g^2_{ch}}{4\pi}\frac{m^2_{\pi}}{4m^2_u}
 \nonumber \\
V^{\sigma,LS}_{ij} & = & -\frac{\alpha_{ch}}{2m_{\pi}^2}
\frac{\Lambda^2}{\Lambda^2-m_{\sigma}^2}m^3_{\sigma} \left[
G(m_\sigma r_{ij})- \frac{\Lambda^3}{m_{\sigma}^3}G(\Lambda
r_{ij}) \right] [{\mathbf r}_{ij} \times ({\mathbf p}_i-{\mathbf
p}_j)] \cdot({\mathbf \sigma}_i+{\mathbf \sigma}_j),
\nonumber \\
V^{C}(r_{ij})&=& -a_c {\mathbf \lambda}_i \cdot {\mathbf
\lambda}_j (r^2_{ij}+V_0)+V^{C,LS}_{ij}, \nonumber
\\
V^{C,LS}_{ij} & = & -a_c {\mathbf \lambda}_i \cdot{\mathbf
\lambda}_j
\frac{1}{8m_q^2}\frac{1}{r_{ij}}\frac{dV^c}{dr_{ij}}[{\mathbf
r}_{ij} \times ({\mathbf p}_i-{\mathbf p}_j)] \cdot({\mathbf
\sigma}_i+{\mathbf \sigma}_j),~~~~~~ V^{c}=r^{2}_{ij},
\nonumber \\
S_{ij} & = &  \frac{{\mathbf (\sigma}_i \cdot {\mathbf r}_{ij})
({\mathbf \sigma}_j \cdot {\mathbf
r}_{ij})}{r_{ij}^2}-\frac{1}{3}~{\mathbf \sigma}_i \cdot {\mathbf
\sigma}_j. \nonumber
\end{eqnarray}
\end{widetext}
Where $S_{ij}$ is quark tensor operator, $Y(x)$, $H(x)$ and $G(x)$
are standard Yukawa functions~\cite{QM}, $T_c$ is the kinetic energy
of the center of mass, $\alpha_{ch} $ is the chiral coupling
constant, determined as usual from the $\pi$-nucleon coupling
constant. All other symbols have their usual meanings. The
parameters of this ChQM Hamiltonian are given in Table
\ref{parameters}.

\subsection{Quark delocalization color screening model}
The model and its extension were discussed in detail in
Ref.\cite{QDCSM1,QDCSM2}. Its Hamiltonian has the same form as
Eq.(1), but without $\sigma$ meson exchange and a phenomenological
color screening confinement potential is used,
\begin{eqnarray}
V^{C}(r_{ij})&=& -a_c {\mathbf \lambda}_i \cdot {\mathbf
\lambda}_j [f(r_{ij})+V_0]+V^{C,LS}_{ij}, \nonumber
\\
 f(r_{ij}) & = &  \left\{ \begin{array}{ll}
 r_{ij}^2 &
 \qquad \mbox{if }i,j\mbox{ occur in the same } \\
 & \qquad \mbox{baryon orbit}, \\
 \frac{1 - e^{-\mu r_{ij}^2} }{\mu} & \qquad
 \mbox{if }i,j\mbox{ occur in different} \\
 & \qquad \mbox{baryon orbits}.
 \end{array} \right.
\end{eqnarray}
Here, $\mu$ is the color screening constant to be determined by
fitting the deuteron mass in this model.
The quark delocalization in QDCSM is realized by allowing the single
particle orbital wave function of QDCSM as a linear combination of
left and right Gaussian, the single particle orbital wave functions
in the ordinary quark cluster model,
\begin{eqnarray}
\psi_{\alpha}(\vec{S}_i ,\epsilon) & = & \left(
\phi_{\alpha}(\vec{S}_i)
+ \epsilon \phi_{\alpha}(-\vec{S}_i)\right) /N(\epsilon), \nonumber \\
\psi_{\beta}(-\vec{S}_i ,\epsilon) & = &
\left(\phi_{\beta}(-\vec{S}_i)
+ \epsilon \phi_{\beta}(\vec{S}_i)\right) /N(\epsilon), \nonumber \\
N(\epsilon) & = & \sqrt{1+\epsilon^2+2\epsilon e^{-S_i^2/4b^2}}. \label{1q} \\
\phi_{\alpha}(\vec{S}_i) & = & \left( \frac{1}{\pi b^2}
\right)^{3/4}
   e^{-\frac{1}{2b^2} (\vec{r}_{\alpha} - \vec{S}_i/2)^2} \nonumber \\
\phi_{\beta}(-\vec{S}_i) & = & \left( \frac{1}{\pi b^2}
\right)^{3/4}
   e^{-\frac{1}{2b^2} (\vec{r}_{\beta} + \vec{S}_i/2)^2}. \nonumber
\end{eqnarray}
The mixing parameter $\epsilon(S)$ is not an adjusted one but
determined variationally by the dynamics of the multi-quark system
itself. This assumption allows the multi-quark system to choose its
favorable configuration in the interacting process. It has been used
to explain the cross-over transition between hadron phase and
quark-gluon plasma phase~\cite{liu}. The model parameters are fixed
as follows: The $u,d$-quark mass difference is neglected and
$m_u$=$m_d$ is assumed to be exactly $1/3$ of the nucleon mass,
namely $m_u$=$m_d$=$313$ MeV. The $\pi$ mass takes the experimental
value. The $\Lambda$ takes the same values as in Ref.\cite{QM},
namely $\Lambda$=4.2 fm$^{-1}$. The chiral coupling constant
$\alpha_{ch}$ is determined from the $\pi NN$ coupling constant as
usual. The other parameters b, $a_c$, $V_0$, and $\alpha_s$ are
determined by fitting the nucleon and $\Delta$ masses and the
stability of nucleon size b with the variation of quark mass m. All
parameters used are listed in Table \ref{parameters}. In order to
compare the intermediate-range attraction mechanism, the $\sigma$
meson exchange in ChQM and quark delocalization and color screening
in QDCSM, the same values of parameters:
$b,~\alpha_s,~\alpha_{ch},~m_u,~m_\pi,~\Lambda$ are used for these
two models. Thus, these two models have exactly the same
contributions from one-gluon-exchange and $\pi$ exchange. The only
difference of the two models is coming from the short and
intermediate-range part, $\sigma$ exchange for ChQM, quark
delocalization and color screening for QDCSM. To show the
sensitivity of the QDCSM to the model parameters, the results of
another set of model parameters (QDCSM2) is also reported.
\begin{table}[ht]
\caption{Parameters of three quark models discussed in this paper.}
\begin{tabular}{lcccc}
\hline
 & {\rm ChQM} & {\rm QDCSM1} & {\rm QDCSM2} & {\rm QDCCM}    \\
\hline\hline
$m_{u,d}({\rm MeV})$        &  313    &  313    & 313    & 313  \\
$b ({\rm fm})$              &  0.518  &  0.518  & 0.60   & 0.518 \\
$a_c({\rm MeV\,fm}^{-2})$   &  46.938 & 56.755  & 18.5   & 56.755 \\
$V_0({\rm fm}^{2})$         &  -1.297 & -0.5279 & -1.3598& -0.5279\\
$\mu ({\rm fm}^{-2})$       &         &  0.45   & 1.00   &      \\
$\alpha_s$                  &  0.485  &  0.485  & 0.996  & 0.485\\
$m_\pi({\rm MeV})$          &  138    &  138    & 138    & 138  \\
$\alpha_{ch}$               &  0.027  & 0.027   & 0.027  & 0.027\\
$m_\sigma ({\rm MeV})$      &  675    &         &        &      \\
$\Lambda ({\rm fm}^{-1})$   &  4.2    &  4.2    &  4.2   &  4.2 \\
\hline
\end{tabular}
\label{parameters}
\end{table}

\subsection{Quark delocalization model with hidden color channels coupling (QDCCM)}

This approach is focused on the hidden color channel effect which
has been ignored almost in all quark model calculations but
certainly should exist in a description based on the fundamental
quark-gluon degree of freedom. In the lattice QCD calculation of
$NN$ interaction~\cite{lqcd} these hidden color channels should have
been included implicitly. However their effect has not yet been
separated. We assume a Hamiltonian which is the same as that of
QDCSM except that the usual quadratic confinement
\begin{equation}
V^{C}(r_{ij})= -k a_c {\mathbf \lambda}_i \cdot {\mathbf
\lambda}_j (r^2_{ij}+V_0).
\end{equation}
\begin{table}[ht]
\caption{The channels used in $NN$ scattering calculations and the
factors $k_1,k_2$ (for recipes 1,2) for each channel (I=1).}
\begin{tabular}{c|lc}
\hline
 $J$ & \mbox{~~~~~~~}\hspace{1in} channels & $k_1/k_2$   \\ \hline
0 & $^1S_0:
NN,\Delta\Delta,~^2\Delta_8~^2\Delta_8,~^4N_8~^4N_8,~^2N_8~^2\Delta_8$, & \\
 & ~~~~~~~~$^2N_8~^2N_8$ & $1.42/1.39$  \\
 & $^5D_0: N\Delta,\Delta\Delta,~^4N_8~^2\Delta_8,~^4N_8~^4N_8,~^4N_8~^2N_8$ &
  \\ \cline{2-3}
& $^3P_0:
NN,N\Delta,\Delta\Delta,~^2\Delta_8~^2\Delta_8,~^4N_8~^2\Delta_8$, & $1.10/1.10$ \\
 & ~~~~~~~~$^4N_8~^4N_8,~^4N_8~^2N_8,~^2N_8~^2\Delta_8,~^2N_8~^2N_8$ &  \\  \hline
1 & $^3P_1:
NN,N\Delta,\Delta\Delta,~^2\Delta_8~^2\Delta_8,~^4N_8~^2\Delta_8$, & $1.35/1.28$  \\
 & ~~~~~~~~$^4N_8~^4N_8,~^4N_8~^2N_8,~^2N_8~^2\Delta_8,~^2N_8~^2N_8$ &  \\  \hline
2 & $^1D_2:
NN,\Delta\Delta,~^2\Delta_8~^2\Delta_8,~^4N_8~^4N_8,~^2N_8~^2\Delta_8$, &   \\
 & ~~~~~~~~$^2N_8~^2N_8$  & $2.00/1.85$ \\
 & $^5S_2(^5D_2): N\Delta,\Delta\Delta,~^4N_8~^2\Delta_8,~^4N_8~^4N_8$, & \\
 & ~~~~~~~~~~~~~~~$^4N_8~^2N_8$ &  \\ \cline{2-3}
& $^3P_2:
NN,N\Delta,\Delta\Delta,~^2\Delta_8~^2\Delta_8,~^4N_8~^2\Delta_8$, & $1.75/1.66$  \\
 & ~~~~~~~~$^4N_8~^4N_8,~^4N_8~^2N_8,~^2N_8~^2\Delta_8,~^2N_8~^2N_8$ &  \\ \cline{2-3}
& $^3F_2:
NN,N\Delta,\Delta\Delta,~^2\Delta_8~^2\Delta_8,~^4N_8~^2\Delta_8$, & $1.00/1.00$  \\
 & ~~~~~~~~$^4N_8~^4N_8,~^4N_8~^2N_8,~^2N_8~^2\Delta_8,~^2N_8~^2N_8$ &  \\  \hline
3 & $^3F_3:
NN,N\Delta,\Delta\Delta,~^2\Delta_8~^2\Delta_8,~^4N_8~^2\Delta_8$, & $1.00/1.00$  \\
 & ~~~~~~~$~^4N_8~^4N_8,~^4N_8~^2N_8,~^2N_8~^2\Delta_8,~^2N_8~^2N_8$ & \\  \hline
4 & $^3F_4:
NN,N\Delta,\Delta\Delta,~^2\Delta_8~^2\Delta_8,~^4N_8~^2\Delta_8$, & $1.00/1.00$  \\
 & ~~~~~~~$~^4N_8~^4N_8,~^4N_8~^2N_8,~^2N_8~^2\Delta_8,~^2N_8~^2N_8$ &  \\ \cline{2-3}
 & $^1G_4:
NN,\Delta\Delta,~^2\Delta_8~^2\Delta_8,~^4N_8~^4N_8,~^2N_8~^2\Delta_8,$
& $1.00/1.00$  \\
 & ~~~~~~~~$^2N_8~^2N_8 $ &   \\ \cline{2-3}
& $^3H_4:
NN,N\Delta,\Delta\Delta,~^2\Delta_8~^2\Delta_8,~^4N_8~^2\Delta_8$, & $1.00/1.00$  \\
 & ~~~~~~~~$^4N_8~^4N_8,~^4N_8~^2N_8,~^2N_8~^2\Delta_8,~^2N_8~^2N_8$ &  \\  \hline
 5 & $^3H_5:
NN,N\Delta,\Delta\Delta,~^2\Delta_8~^2\Delta_8,~^4N_8~^2\Delta_8$, & $1.00/1.00$  \\
 & ~~~~~~~~$^4N_8~^4N_8,~^4N_8~^2N_8,~^2N_8~^2\Delta_8,~^2N_8~^2N_8$ &  \\  \hline
 6 & $^3H_6:
NN,N\Delta,\Delta\Delta,~^2\Delta_8~^2\Delta_8,~^4N_8~^2\Delta_8$, & $1.00/1.00$  \\
 & ~~~~~~~~$^4N_8~^4N_8,~^4N_8~^2N_8,~^2N_8~^2\Delta_8,~^2N_8~^2N_8$ & \\ \cline{2-3}
& $^1I_6:
NN,\Delta\Delta,~^2\Delta_8~^2\Delta_8,~^4N_8~^4N_8,~^2N_8~^2\Delta_8,$
& $1.00/1.00$  \\
 & ~~~~~~~$^2N_8~^2N_8 $ &  \\  \hline
\end{tabular}
\end{table}
is used but with an additional multiplying factor $k$. For the
color-singlet channels (two baryon clusters are in the color-singlet
states), the factor $k$ takes the value 1. For the hidden color
channels, two recipes are used. Recipe 1 (QDCCM1): For the coupling
between hidden color channels and the color singlet channels, the
factor $k$ is taken as an adjustable parameter. All the other cases
the factor $k$ is kept 1. Recipe 2 (QDCCM2): The factor $k$ is taken
as adjustable parameter not only for color singlet-hidden color
channels coupling but also for hidden color-hidden color channels.
As for the single quark orbital wave function, the same form Eq.(3)
as that of QDCSM is assumed. This model assumption is inspired by
the lattice QCD calculation : The recent lattice QCD calculations
show that the interactions among quarks are genuinely multi-body
interactions. The color dependent two body confinement interaction
is consistent with the lattice QCD results only for two and three
quark systems in color singlet states but inconsistent with the
multi-body interaction obtained in lattice QCD for multi-quark
systems \cite{lat}. So the direct extension of the color dependent
two body confinement interaction from two- or three-quark system to
multi-quark system as used in the most quark model calculations is
questionable. The calculation based on the direct extension can not
describe the $NN$ scattering quantitatively well even after
including hidden color channels coupling as shown in QDCCM0 might be
an indication of this inadequacy. In fact, for multi-quark systems
and color octet nucleons, quark pairs are not always in color
antisymmetric state but also color symmetric ones. The color factor
{\boldmath$\lambda$}$_{i} \cdot$ {\boldmath$\lambda$}$_{j}$ will
give rise to anti-confinement interaction for symmetric quark pairs
\cite{deconfine}. In QDCSM mentioned above, we used a color
screening confinement interaction to model the effect of this
multi-body confinement interaction obtained in lattice QCD. Here we
study directly the effect of hidden color channel coupling to test
if the phenomenological color screening confinement is an effective
description of the hidden channel coupling. In order to simplify the
numerical calculation, a two body confinement interaction form
Eq.(4) is still assumed but with an additional adjustable
multiplying factor aimed to reflect the effect of the lattice QCD
multi body confinement. At the same time, the model parameters are
kept to the same as those of QDCSM1, except the color screening
confinement form Eq.(2) is replaced by the usual quadratic
confinement form Eq.(4). This is aimed to let the effect of hidden
color channel coupling stand out.
\begin{table}[h]
\caption{The channels used in $NN$ scattering calculations and the
factors $k_1,k_2$ (for recipes 1,2) for each channel (I=0).}
\begin{tabular}{c|lc}
\hline
 $J$ & \mbox{~~~~~~~}\hspace{1in} channels & ~$k_1/k_2$~   \\ \hline
 1 & $^3S_1(^3D_1): NN,\Delta\Delta, ~^2\Delta_8~ ^2\Delta_8,
 ^4N_8 ~^4N_8$, & \\
 & ~~~~~~~~~~~~~~~$^4N_8 ~^2N_8,~^2N_8 ~^2N_8
 $ & 1.40/1.38  \\
  & $^7D_1: \Delta\Delta,~^4N_8~^4N_8$ &   \\  \cline{2-3}
& $^1P_1:
NN,\Delta\Delta,~^2\Delta_8~^2\Delta_8,~^4N_8~^4N_8,~^2N_8~^2N_8$
& $1.80/1.70$  \\
 & $^5P_1: \Delta\Delta,~^4N_8 ~^4N_8, ~^4N_8~^2N_8$ &  \\  \hline
2 & $^3D_2:
NN,\Delta\Delta,~^2\Delta_8~^2\Delta_8,~^4N_8~^4N_8,~^4N_8~^2N_8$, & \\
 & ~~~~~~~~~$^2N_8~^2N_8$
& $1.00/1.00$  \\
 & $^7D_2: \Delta\Delta,~^4N_8~^4N_8$ &  \\ \hline
 3 & $^3D_3:
NN,\Delta\Delta,~^2\Delta_8~^2\Delta_8,~^4N_8~^4N_8,~^4N_8~^2N_8$, & \\
 & ~~~~~~~~$^2N_8~^2N_8$
& $2.40/2.20$  \\
 & $^7S_3(^7D_3): \Delta\Delta,~^4N_8~^4N_8$ & \\ \cline{2-3}
& $^1F_3: NN,\Delta\Delta,~^2\Delta_8~^2\Delta_8,~^4N_8~^4N_8,$ &
 $1.00/1.00$  \\
 & ~~~~~~~~$^2N_8~^2N_8$ &  \\ \cline{2-3}
& $^3G_3: NN,\Delta\Delta,~^2\Delta_8~^2\Delta_8,~^4N_8~^4N_8,$ &
$1.00/1.00$  \\
 & ~~~~~~~~$^4N_8~^2N_8,~^2N_8~^2N_8$ &  \\ \hline
 4 & $^3G_4:
NN,\Delta\Delta,~^2\Delta_8~^2\Delta_8,~^4N_8~^4N_8,~^4N_8~^2N_8,$
& $1.00/1.00$  \\
 & ~~~~~~~ $^2N_8~^2N_8$ &  \\ \hline
 5 & $^3G_5:
NN,\Delta\Delta,~^2\Delta_8~^2\Delta_8,~^4N_8~^4N_8,~^4N_8~^2N_8,$
& $1.00/1.00$  \\
 & ~~~~~~~ $^2N_8~^2N_8$ &  \\ \cline{2-3}
& $^1H_5: NN,\Delta\Delta,~^2\Delta_8~^2\Delta_8,~^4N_8~^4N_8,$ &
 $1.00/1.00$  \\
 & ~~~~~~~ $^2N_8~^2N_8$ &   \\ \cline{2-3}
 & $^3I_5:
NN,\Delta\Delta,~^2\Delta_8~^2\Delta_8,~^4N_8~^4N_8,~^4N_8~^2N_8,$
& $1.00/1.00$  \\
 & ~~~~~~ $^2N_8~^2N_8$ &  \\ \hline
 6 & $^3I_6:
NN,\Delta\Delta,~^2\Delta_8~^2\Delta_8,~^4N_8~^4N_8,~^4N_8~^2N_8,$
& $1.00/1.00$  \\
 & ~~~~~~ $^2N_8~^2N_8$ &   \\ \hline
 7 & $^3I_7:
NN,\Delta\Delta,~^2\Delta_8~^2\Delta_8,~^4N_8~^4N_8,~^4N_8~^2N_8,$
& $1.00/1.00$  \\
 & ~~~~~~ $^2N_8~^2N_8$ &  \\ \hline
\end{tabular}
\end{table}

\setcounter{figure}{1}
\section{The results and discussions}
\begin{figure*}
\epsfxsize=7.0in \epsfbox{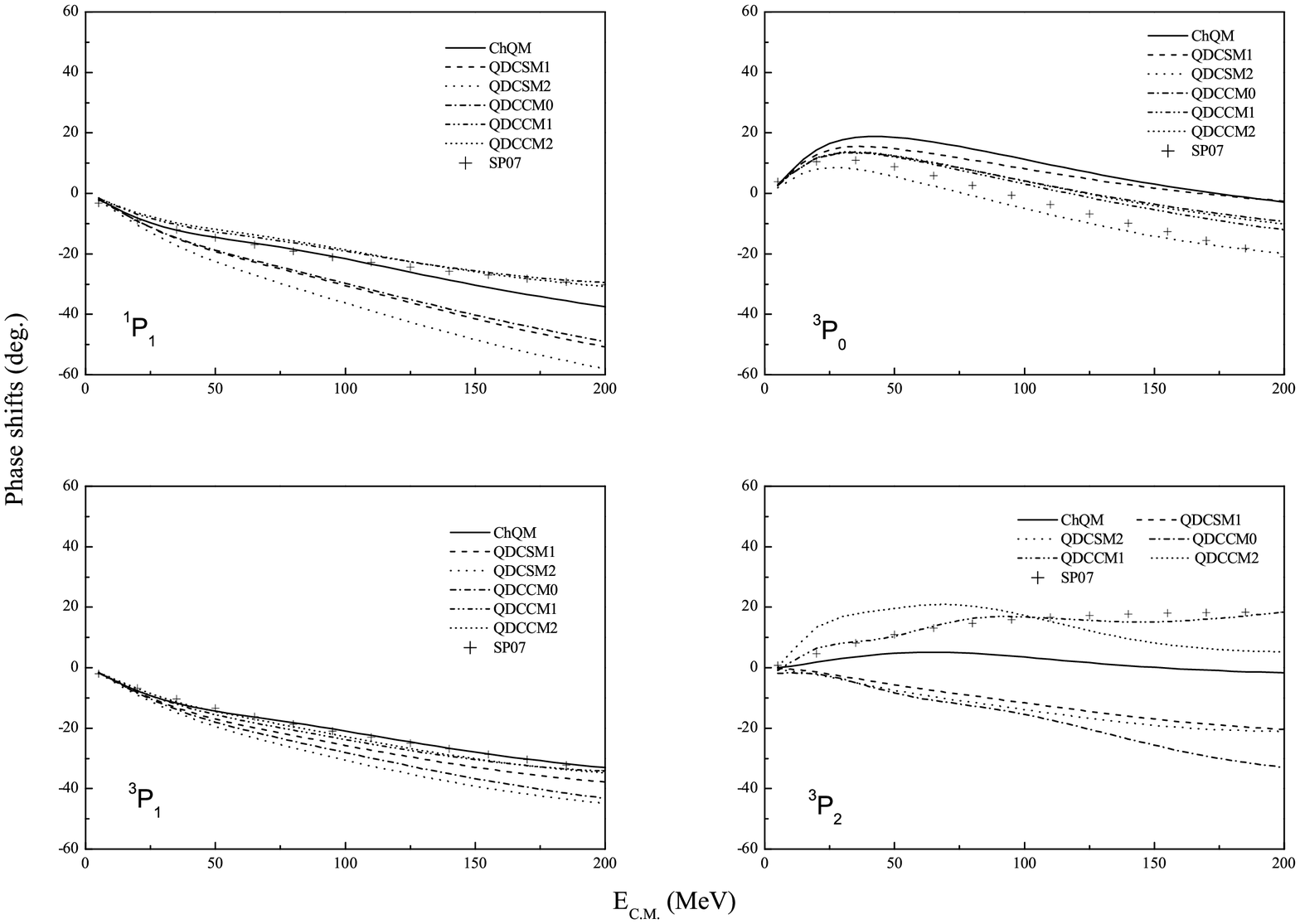}
\vspace{-0.4in}
\caption{The phase shifts of $NN$ $P$ wave scattering.}
\end{figure*}

We calculated the $NN$ scattering phase shifts of different partial
waves ($S$, $P$, $D$, $F$, $G$, $H$ and $I$ waves) by three quark
models mentioned above. To look for non-strange dibaryon resonances,
a systematic calculation of $NN$ scattering phase shifts with
explicit coupling to $N\Delta$ and $\Delta\Delta$ channels is also
done. The resonating-group method (RGM), described in more detail in
\begin{center}
\epsfxsize=3.5in \epsfbox{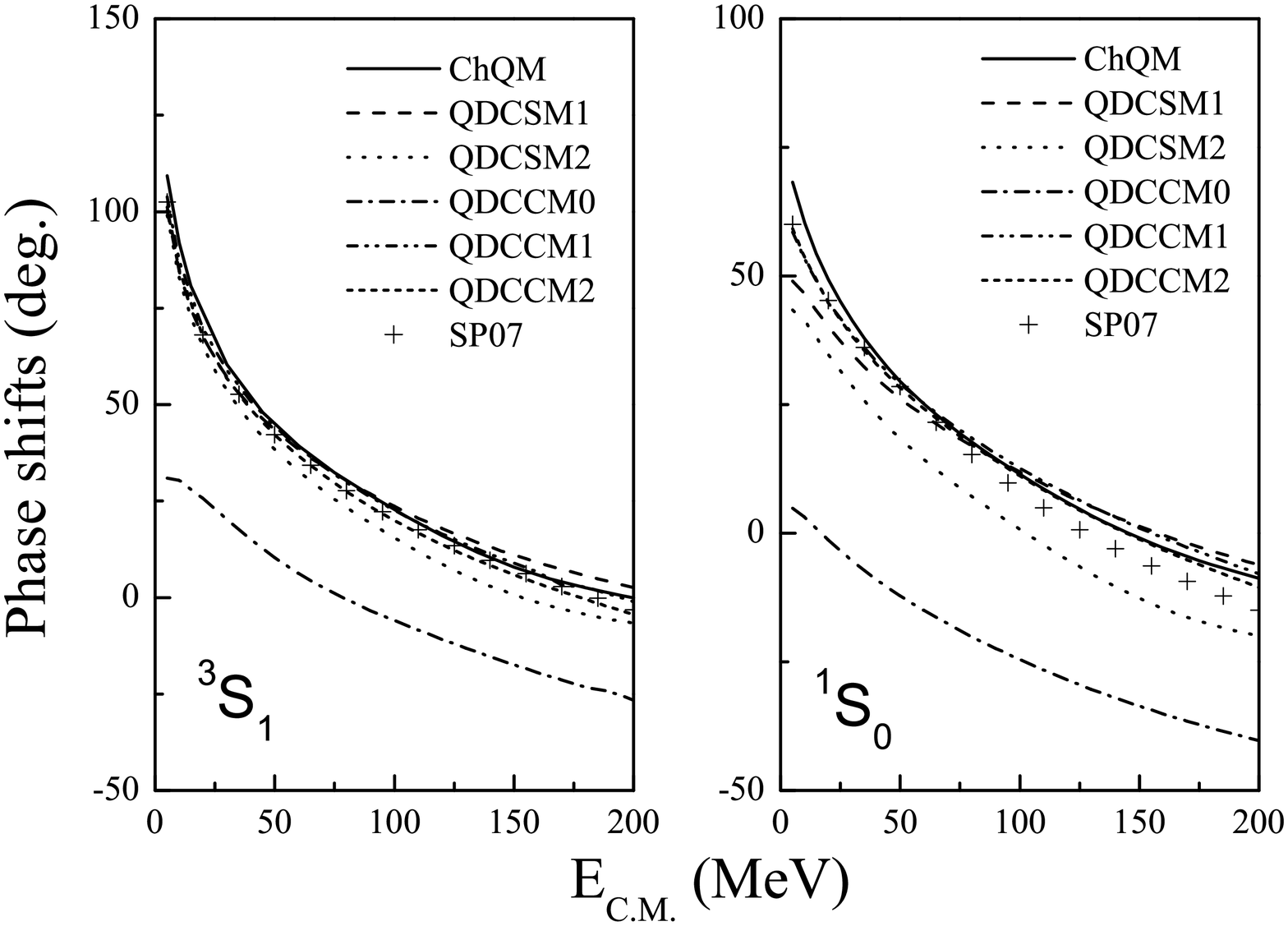}

FIG 1. The phase shifts of $NN$ $S$-wave scattering.
\end{center}
Ref.\cite{RGM}, is used to do the calculation. The experimental
information used for the comparison is the partial-wave solution
SP07 \cite{SP07} of $NN$ scattering data. For QDCSM, the color
screening parameter $\mu$ is fixed by deuteron properties and no
other parameters readjusted. For the third approach (QDCCM), the
channels included in different partial waves are listed in Table II and III.
The multiplying factors $k_1,k_2$ are adjusted to fit the $NN$ phase
shifts of SP07. The calculated results for $NN$ scattering phase
shifts are presented in section A; deuteron properties are shown in
section B and the discussions of the dibaryon resonances are given
in section C.

\subsection{$NN$ scattering phase shifts}

(1) $S-$waves: Fig. 1 shows the $NN$ scattering phase shifts for
$^{3}S_{1}$ and $^{1}S_{0}$ partial waves. A perfect fit is obtained
for both ChQM and QDCSM1 (QDCSM2 gives a little less attraction).
The dominant contribution to the $S$-wave phase shift comes from the
central part of the potentials. The agreement between two models
means these two quark models give the same $NN$ interaction, at
least the same central part. For QDCCM, QDCCM1 and QDCCM2 also give
good descriptions of $NN$ $^{3}S_{1}$ and $^{1}S_{0}$ scattering
phase shifts by including the hidden color channels and adjusting
the color confinement interaction strength, while with the usual
color confinement interaction strength ($k=1$), the model (QDCCM0)
calculated phase shifts are far from the measured ones.

(2) $P-$waves: Fig. 2 shows the $NN$ scattering phase shifts of
$^{1}P_{1}$, $^{3}P_{0}$, $^{3}P_{1}$ and $^{3}P_{1}$ partial waves.
For $^{1}P_{1}$ and $^{3}P_{1}$, ChQM and QDCCM gave an almost
perfect description of the experimental data. The $^{1}P_{1}$ phase
shift is mainly determined by the central repulsion. The theoretical
phase shifts of QDCSM and QDCCM0 are lower than experimental ones
which show that these two models give a too strong repulsion. For
$^{3}P_{0}$, QDCSM2 described the experimental data better than
others. For QDCCM we do not have to adjust the color confinement
interaction strength $a_c$ too much ($k= 1.1$ for both QDCCM1 and
QDCCM2), so both QDCCM1 and QDCCM2, even QDCCM0, can fit the
$^{3}P_{0}$ phase shifts reasonable well. ChQM and QDCSM1 give too
strong attraction. For $^{3}P_{2}$, QDCCM1 gives a perfect fit.
ChQM, QDCSM and QDCCM0 do not have enough attraction. Fig.~3 shows
central, spin-orbit and tensor components of the $^3P_J$ phase
shifts. Clearly, ChQM, QDCSM and QDCCM0 do not give strong enough
attraction in the central and spin-orbit parts. In OBE,
$\pi\rho,~\pi\omega$-exchange, which might have not been reproduced
in the quark model calculations, are also needed to reproduce the
$P$-wave phase shifts~\cite{OBE}.
\begin{figure}
\epsfxsize=3.5in \epsfbox{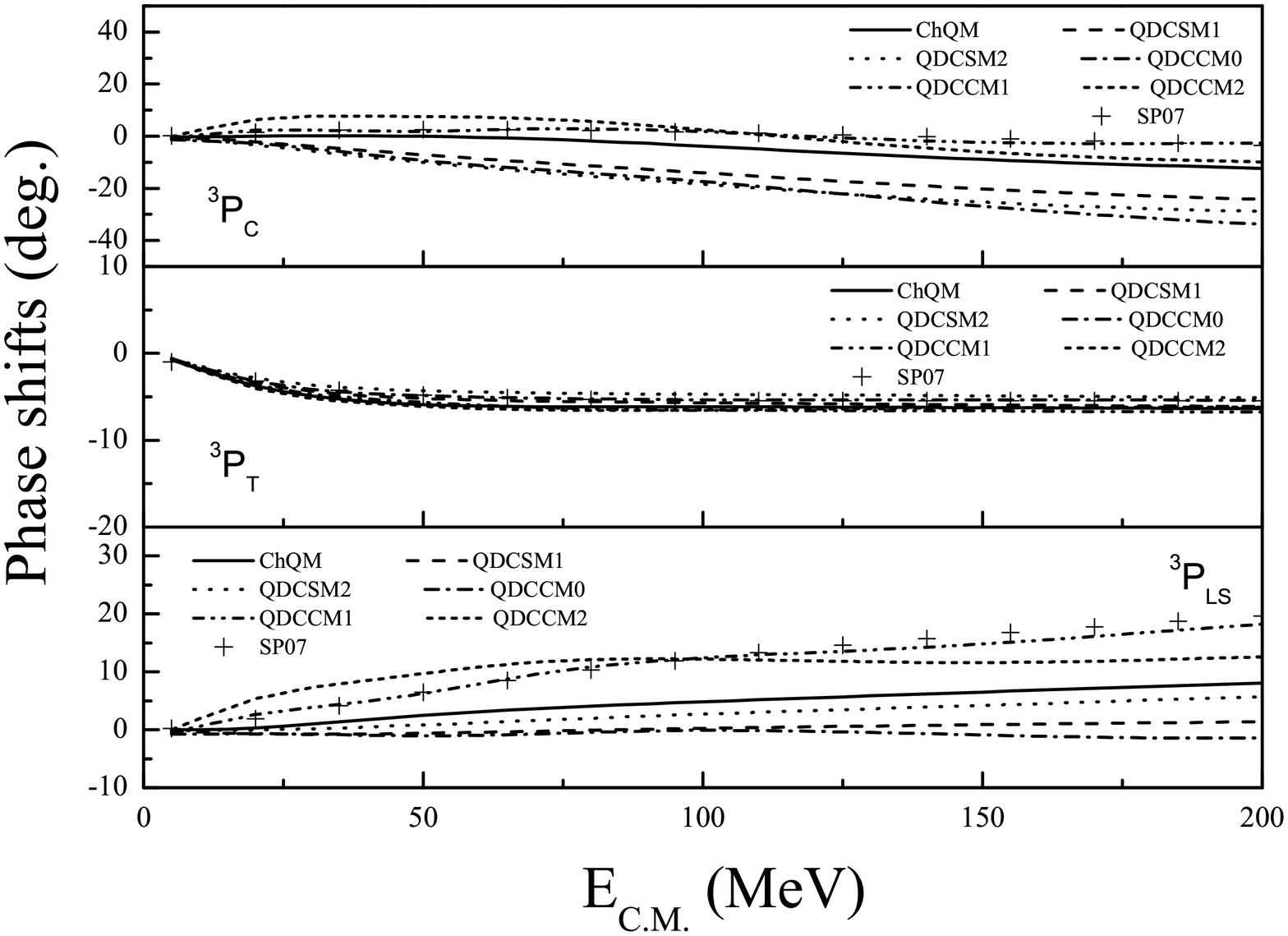}
\caption{Central, spin-orbit and tensor components of $NN$ $P$ wave scattering.}
\end{figure}

(3) $D-$waves: Fig. 4 shows the $NN$ scattering phase shifts of
$^{3}D_{1}$, $^{3}D_{2}$, $^{3}D_{3}$ and $^{1}D_{2}$ partial waves.
For $^3D_1$, all the models fit the experimental data well except
QDCCM0. For $^3D_2$, QDCSM1 and QDCSM2 give a very good description
of the experimental data (QDCSM2 is a little better), ChQM gives too
strong attraction. For QDCCM, we find that we do not need adjust the
color confinement interaction strength for this channel, QDCCM0 can
fit the experimental scattering phase shifts. For $^{3}D_{3}$ and
$^{1}D_{2}$, ChQM described the experimental data better than QDCSM.
For QDCCM, both adjusting recipes can give a perfect fit to the
experimental data.
\begin{figure*}[p]
\epsfxsize=6.5in \epsfbox{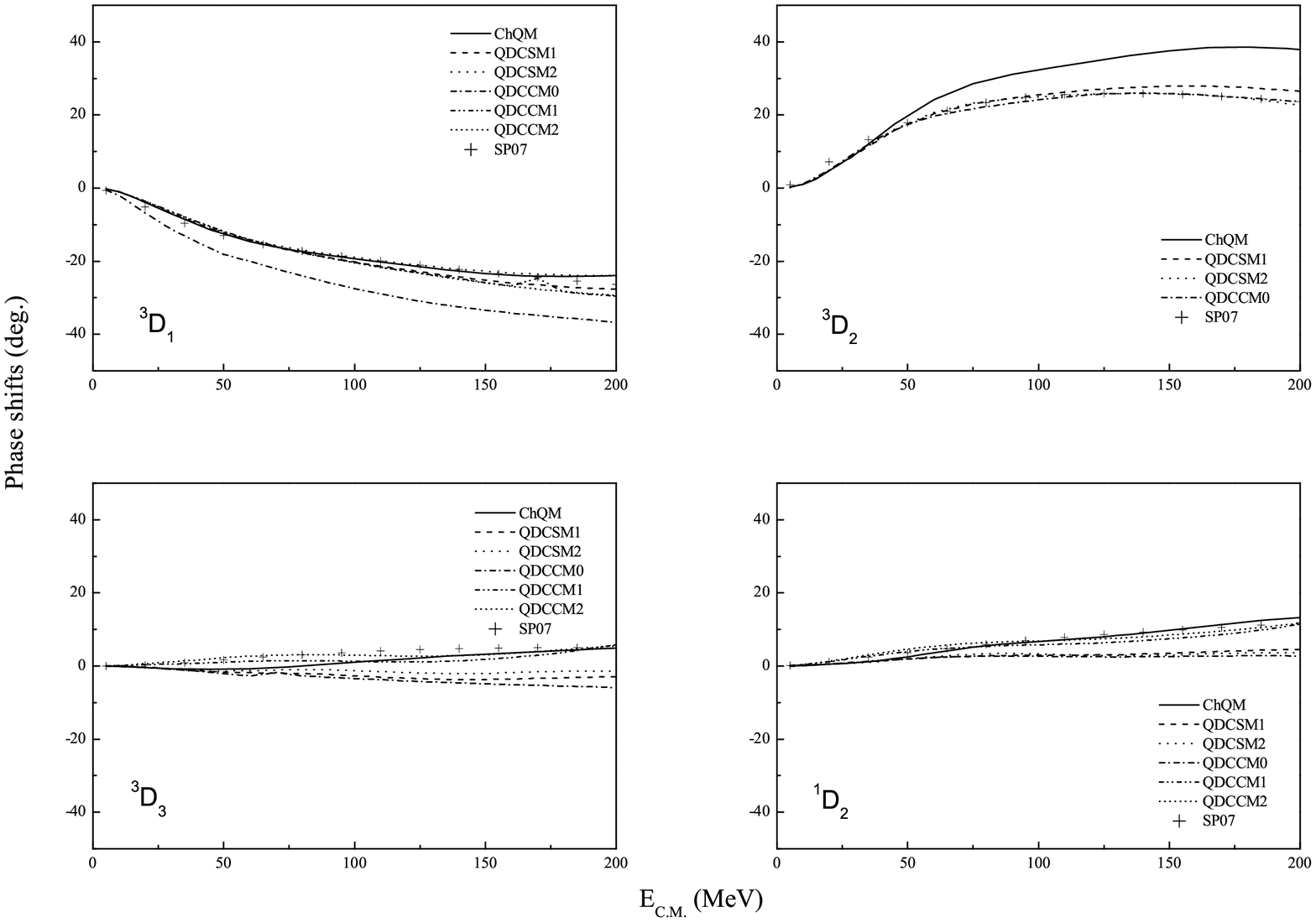}
\vspace{-0.4in}
\caption{The phase shifts of $NN$ $D$ wave scattering.}
\end{figure*}
\begin{figure*}[p]
\epsfxsize=6.5in \epsfbox{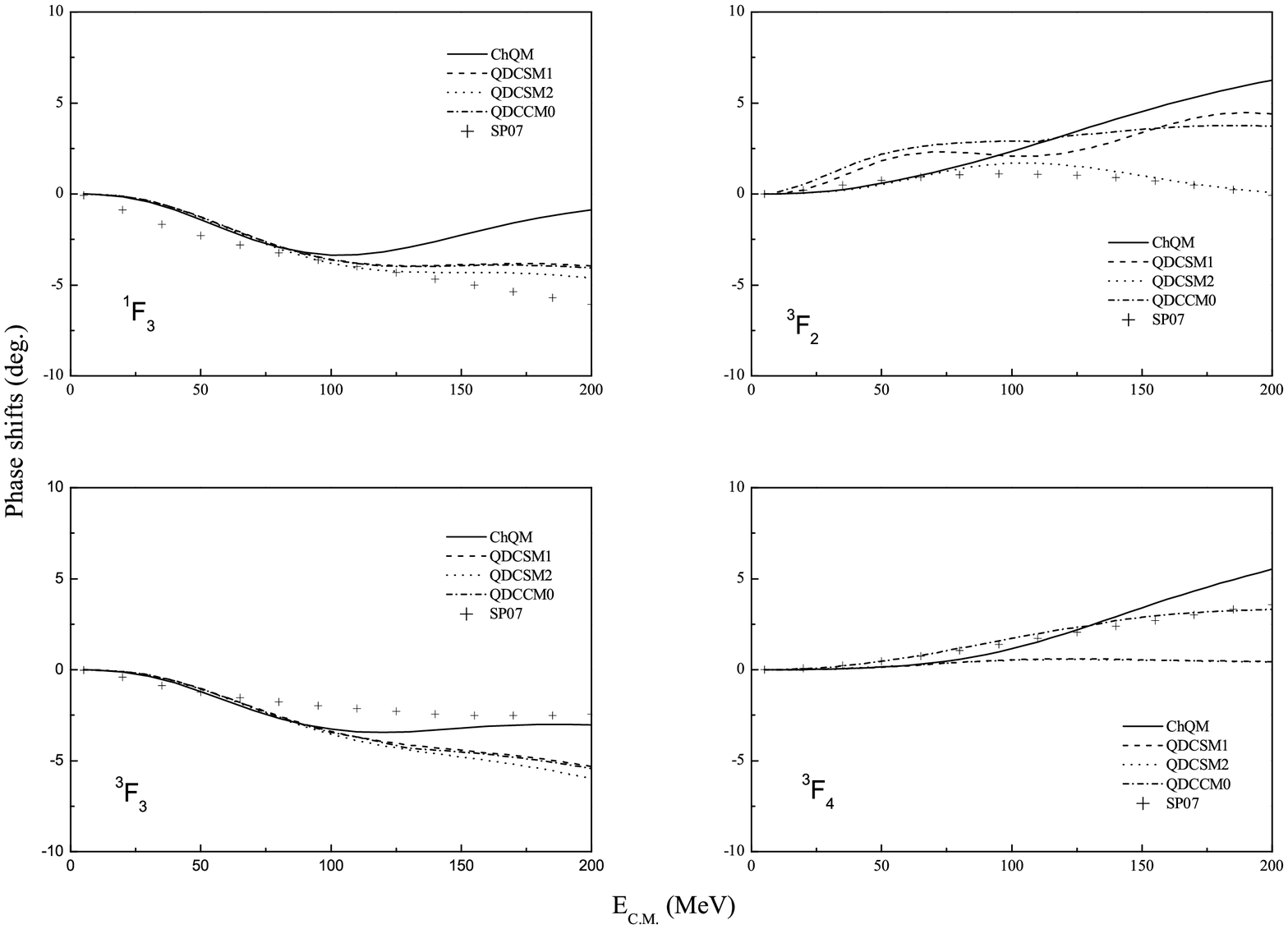}
\vspace{-0.4in}
\caption{The phase shifts of $NN$ $F$ wave scattering.}
\end{figure*}
\begin{figure*}[p]
\epsfxsize=6.5in \epsfbox{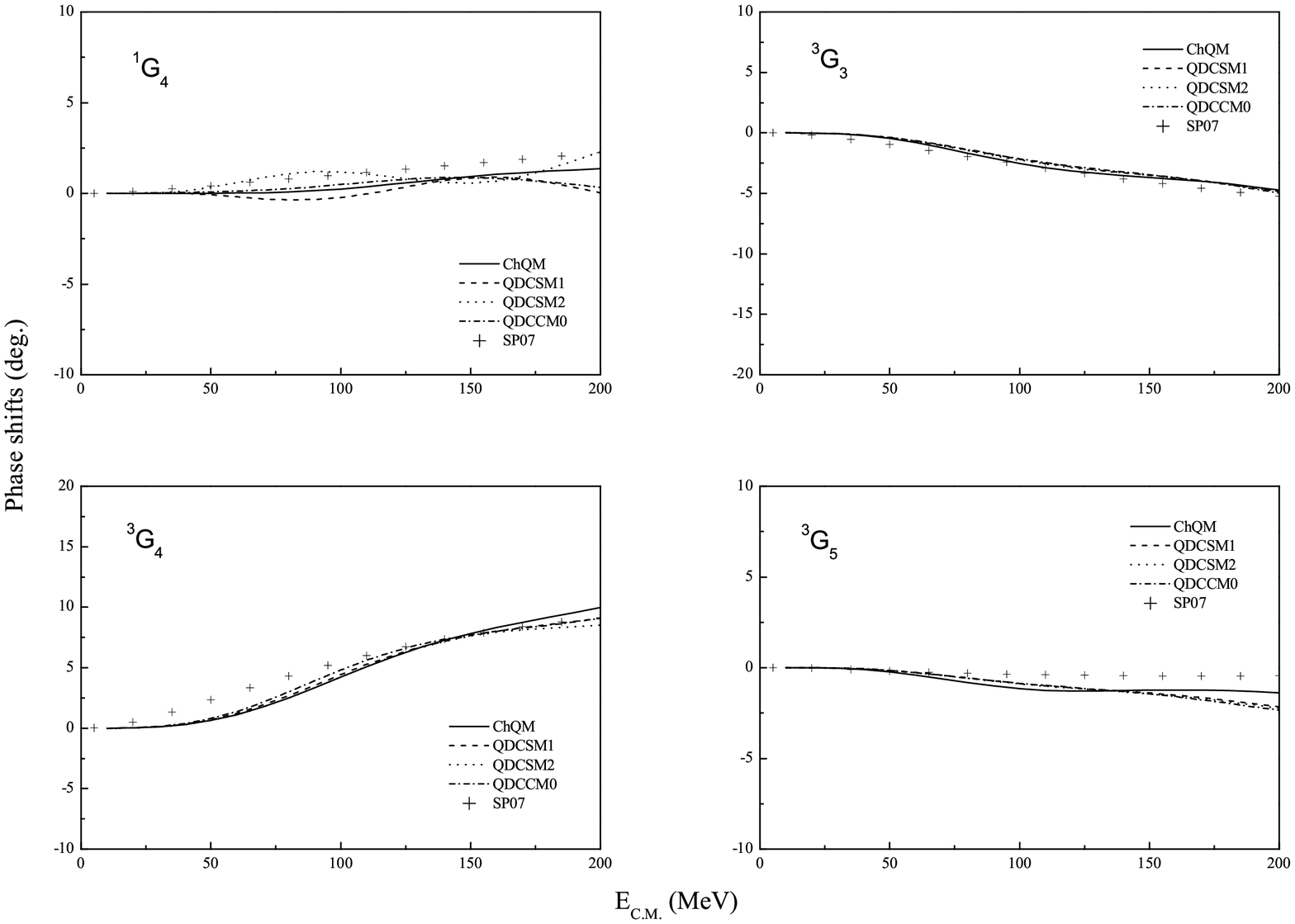}
\vspace{-0.4in}
\caption{The phase shifts of $NN$ $G$ wave scattering.}
\end{figure*}
\begin{figure*}[p]
\epsfxsize=6.5in \epsfbox{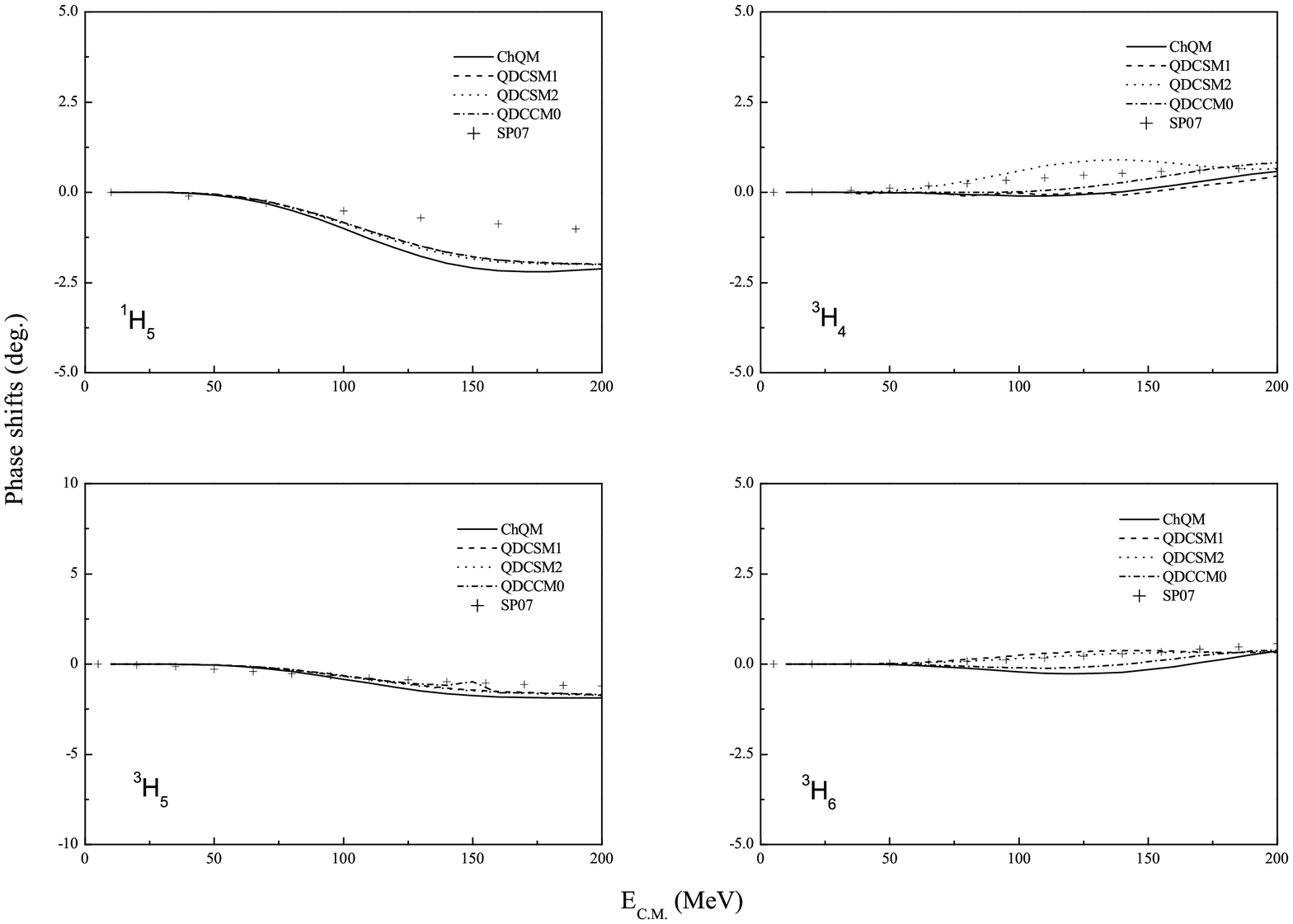}
\vspace{-0.4in}
\caption{The phase shifts of $NN$ $H$ wave scattering.}
\end{figure*}
\begin{figure*}[t]
\epsfxsize=6.5in \epsfbox{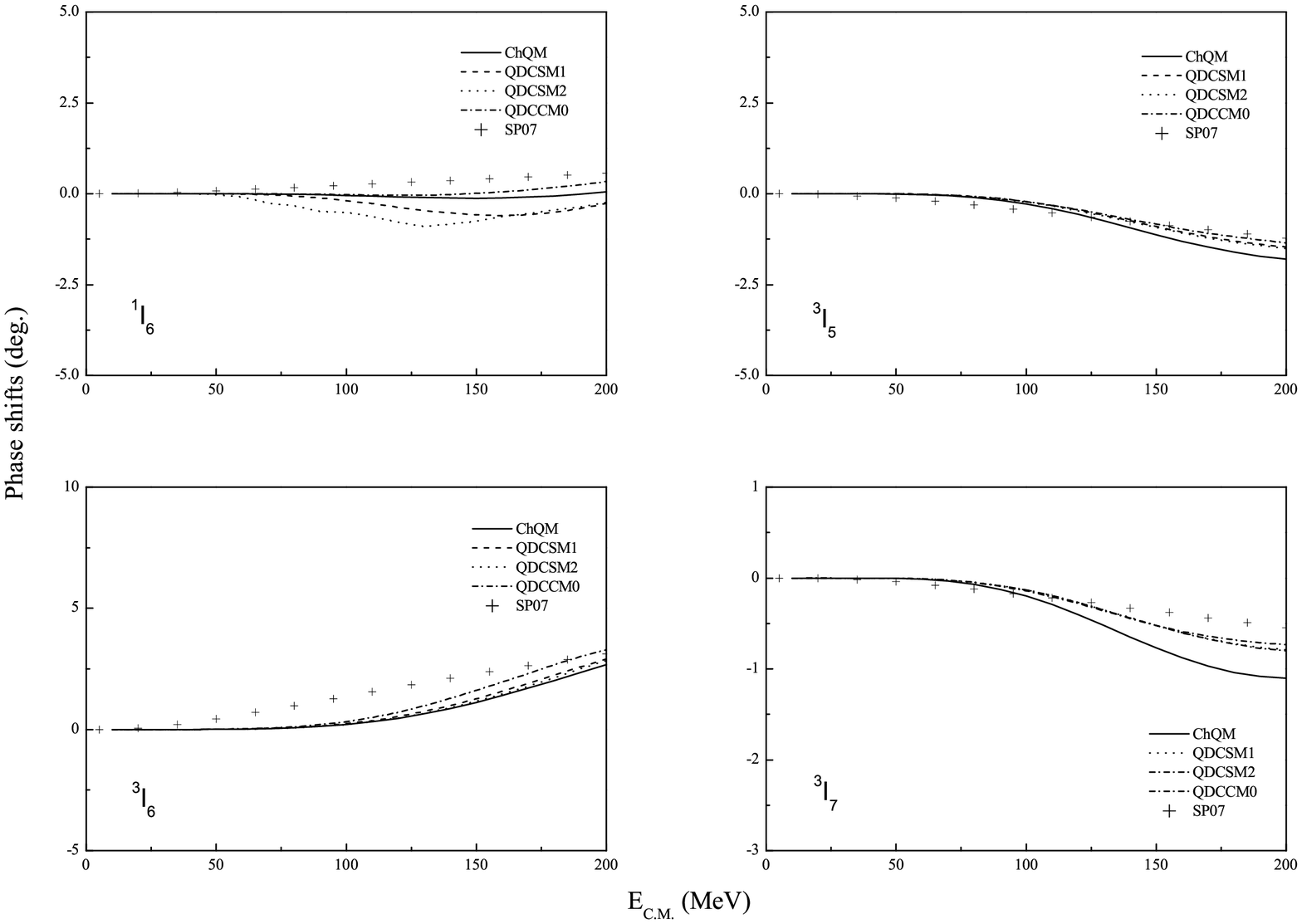}
\vspace{-0.4in}
\caption{The phase shifts of $NN$ $I$ wave scattering.}
\end{figure*}

(4) $F$-wave: The calculated $^{1}F_{3}$, $^{3}F_{2}$, $^{3}F_{3}$
and $^{3}F_{4}$ $NN$ phase shifts are shown in Fig. 5. For $F$ wave
scattering, we find that QDCCM0 already fit the experimental
scattering phase shifts reasonably so we did not fine tune the
confinement strength . All the models give a good description of the
experimental data reasonably in the low energy region ($E_{c.m.} <
100$ MeV). Above 100 MeV, the model predictions deviate more or less
from the experimental data. For $^{3}F_{2}$ QDCSM2 gave much better
fit to the experimental data than QDCSM1, ChQM and QDCCM0. For
$^{1}F_{3}$, QDCSM1, QDCSM2 and QDCCM0 all give better fit to the
experimental data than ChQM, especially at higher energy. However,
for $^{3}F_{3}$, ChQM is closer to the experimental data than other
models. For $^{3}F_{4}$, a perfect fit is obtained for QDCCM0, ChQM
has a little too strong attraction at high energy and QDCSM gives a
too weak attraction.

(5) $G$-wave: The $NN$ phase shifts of $^{3}G_{3}$, $^{1}G_{4}$,
$^{3}G_{4}$ and $^{3}G_{5}$ are shown in Fig. 6. All the models can
describe the experimental data. We do not have to adjust the color
confinement interaction strength for QDCCM here.

(6) $H$-wave: Fig. 7 shows the calculated $^{1}H_{5}$,
$^{3}H_{4}$, $^{3}H_{5}$ and $^{3}H_{6}$ $NN$ phase shifts. For
$H$-wave phase shifts, all models fit to the
experimental data equally well.
We also find that we do not have to adjust the
color confinement interaction strength for QDCCM here.

(7) $I$-wave: The calculated $^{3}I_{5}$, $^{1}I_{6}$, $^{3}I_{6}$
and $^{3}I_{7}$ $NN$ phase shifts are shown in Fig. 8. For
$I$-wave phase shifts, all the models give almost
the same results and fit the experimental data well. Again
the color confinement interaction strength for QDCCM do not need
to be adjusted here.

For high $L$ partial waves, the long range $\pi$ exchange dominates
the interaction. Three quark models have the same $\pi$ exchange and
therefore they give almost the same results for $L\ge{3}$ and we
do not have to adjust the multiplying factor for the QDCCM for these
high $L$ partial wave.

These numerical results (Figs.1-8) show that by including the hidden
color channels and adjusting the color confinement interaction
strength, both adjusting recipes can fit the $NN$ scattering phase
shifts well. From the calculated $S,P,D$-wave phase shifts of $NN$
scattering in QDCCM0, we can see that the attraction is always inadequate
because of the appearance of anti-confinement interaction of symmetric
quark pairs. By increasing the strength of confinement, the attraction
coming from the confinement interaction is strengthened, QDCCM1 and
QDCCM2 can give a good description of the experimental data.
We take these results as an indication that the short
and intermediate range $NN$ interaction is caused by the nucleon
internal structure and its distortion both in orbital and color
spaces in the interacting process. These are quite the same as the
atomic internal structure and its distortion in orbital space which
give rise to the molecular covalent bond. The Anderson's
conjecture~\cite{Anderson} is verified here. The phenomenological
color screening confinement might be an effective description of the
hidden color channel coupling. The phenomenological $\sigma$ meson
exchange used in OBE and ChQM might be an effective description of
the more complicated nucleon distortion in the $NN$ interaction
process as described in QDCSM and QDCCM. This mechanism also gives a
natural explanation why does the $NN$ interaction between two color
singlet nucleons is so similar to the molecular interaction between
two charge neutral atoms except the energy and length scale
difference.

\subsection{Deuteron}

All these three models are used to calculate the properties of
deuteron, the results are shown in Table IV. Both ChQM and QDCSM give
a good description of deuteron. For QDCSM, by adjusting the color
screening parameter, the same results for deuteron can be obtained
for different baryon size $b$. Because of the large separation
between the proton and neutron in the deuteron, the properties of
deuteron mainly reflect the long-range part of the nuclear force.
The same $\pi$-exchange used in the two models assure the properties
of deuteron be fitted equally well. However $\pi$ exchange alone can
not provide strong enough intermediate-range attraction to make the
deuteron bound. In ChQM, it is the phenomenological $\sigma$ meson
exchange which provides the intermediate range attraction. In QDCSM
it is the quark delocalization and color screening which provide the
intermediate range attraction. The fact that both models fit the
deuteron properties well verifies once more that the two
intermediate range attraction mechanisms used in these two models
are equivalent.

Table IV shows that the binding energy and the $D$-wave component of
deuteron can be reproduced(we didn't fine tune the strength of color
confinement to get a better fitting). However the root mean square
radius is too small in comparison to experimental value. This may
indicate that QDCCM with the parameters giving in Table I
gives rise to an $NN$ scattering phase shift equivalent potential
but a little too strong attraction in the short range region, which
tightens up the deuteron.
\begin{table}[ht]
\caption{The properties of deuteron.}
\begin{tabular}{ccccccc}\hline\hline
&  & {\rm ChQM} & {\rm QDCSM1} & {\rm QDCSM2} & {\rm QDCCM1} & {\rm QDCCM2}  \\
\hline
& B (MeV) & 2.0 & 1.94 & 2.01 & 1.0 & 2.2\\
 & $\sqrt{r^2} (fm)$ & 1.96 & 1.93 & 1.94 & 1.2 & 1.1\\
 & $P_D (\%)$ & 4.86 & 5.25 & 5.25 & 4.0 & 4.0\\ \hline\hline
\end{tabular}
\end{table}

\subsection{Dibaryon resonances in $NN$ scattering}
\begin{figure*}
\epsfxsize=6.5in \epsfbox{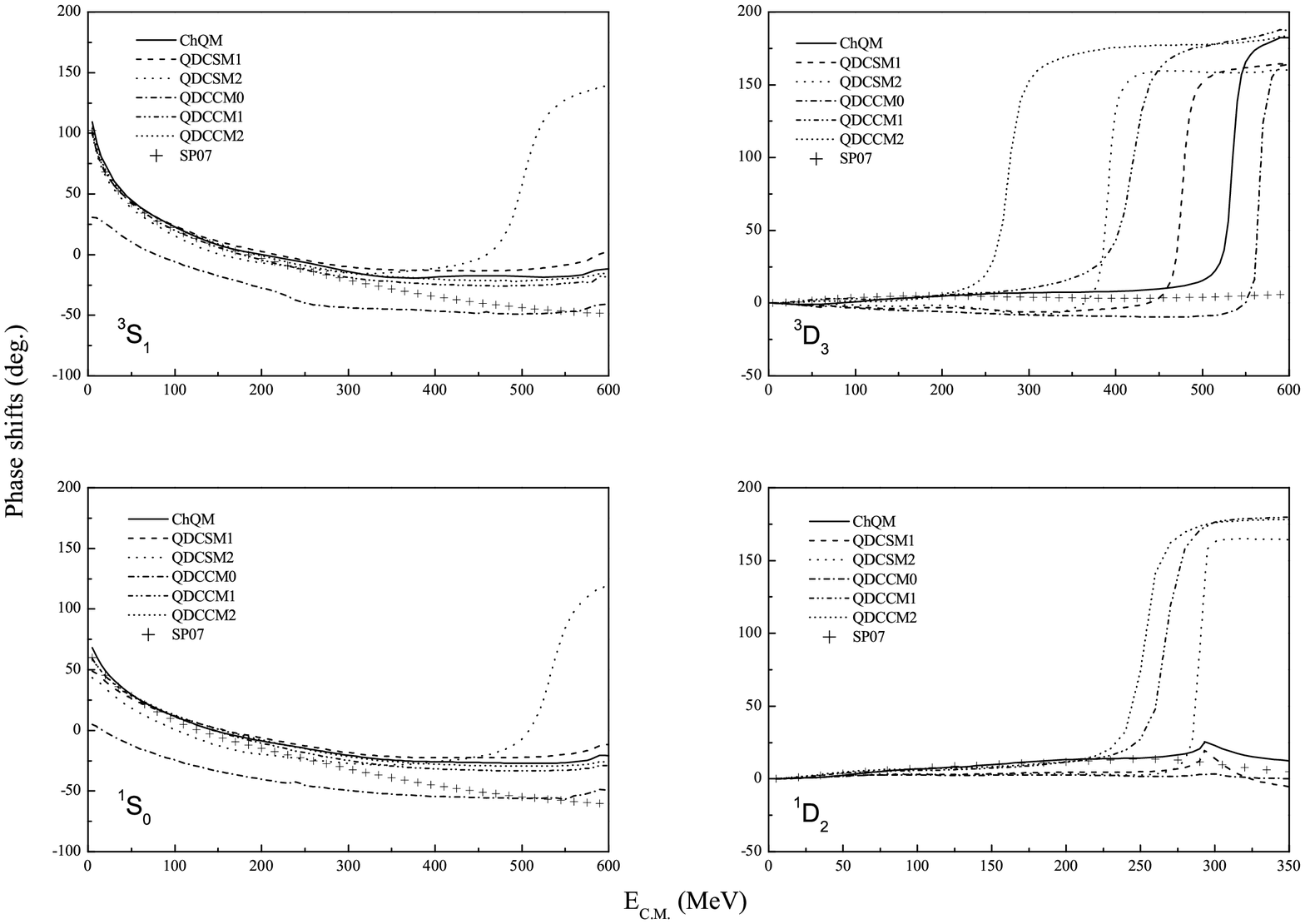}
\vspace{-0.4in}
\caption{The phase shifts of $NN$ $S$ wave and $D$ wave scattering
to energies beyond the $\Delta\Delta$ or $N\Delta$ threshold.}
\end{figure*}

In this part, we show the results of a systematic search for the
possible non-strange dibaryon candidates by three quark models
mentioned above.

The previous calculations~\cite{PRC79} show that there are four
possible dibaryons in the quark model calculations, $N\Delta$ state
with $IJ=12$, $\Delta\Delta$ states with $IJ=01,10,03$. Here the
QDCCM is applied to recalculate these states. All of these dibaryon
states are allowed to decay via the $NN$ channels. In other words,
these dibaryon states appear as resonance states in the $NN$
scattering process. So we calculate the $NN$ scattering phase shifts
by including all the possible channel couplings. The results are
shown in Fig. 9.

(1) $I=0, J=1$: The $^3S_1$ energies of single $\Delta-\Delta$
channel calculation are lower than the corresponding threshold
100-350 MeV in ChQM, QDCCM and QDCSM. The coupling to the
$^{3}S_{1}^{NN}$ channel has an unexpectedly large effect, pushing
up the energy of $^{3}S_{1}^{\Delta-\Delta}$ state $\sim 300$ MeV,
so that only in QDCSM2 it becomes a resonance at 2408 MeV. This very
large mass shift is caused by the central interaction and the
presence of a lower-mass state, the deuteron, in the admixed
$^{3}S_{1}^{NN}$ channel. Mixing with other channels listed in 
Table III, the resonance mass is pushed down a little bit, to 2393 MeV. 
In ChQM and QDCCM, the $^{3}S_{1}$ energies in the single
$\Delta-\Delta$ channel calculation are 100 MeV or more higher than
that in QDCSM2. The additional large mass shift caused by the
coupling to the $NN$ channel then pushes the state above the
$\Delta\Delta$ threshold. So no resonance appears in other models
except QDCSM2. The phase shifts of $^{3}S_{1}^{NN}$ are shown in the
up left corner of Fig. 9, where the phase shifts for $100 < E_{c.m.}
< 400$ MeV from ChQM, QDCSM1, QDCSM2, QDCCM1 and QDCCM2 agree with
each other, as already pointed out in section A. The phase shifts of
$^{3}S_{1}^{NN}$ rises through $\pi/2$ at a resonance mass only in
QDCSM2. So, is there an $IJ = 01$ $\Delta\Delta$ resonance state
with resonance mass $2393$ MeV in the $NN$ $^{3}S_{1}$ scattering
channel is not sure.

(2) $I=0, J=3$: The single $\Delta-\Delta$ channel calculation shows
that the state $^{7}S_{3}^{\Delta\Delta}$ is a bound state in all
models used here. The coupling to the $^{3}D_{3}^{NN}$ channels
causes this bound state change into an elastic resonance. The
resonance mass shift, which is caused by the tensor interaction, is
not large $\sim 3$ MeV. The calculation shows that the mass shift is
always dominated by the $NN$ scattering states below the bound-state
rather than those above it. Coupling to other channels listed in
Table III which are above the $^{7}S_{3}^{\Delta\Delta}$ bound state,
the resonance is pushed down as expected. The calculated
$^{3}D_{3}^{NN}$ phase shifts, shown in the up right corner of Fig. 9, 
rise through $\pi/2$ at the resonance masses in all models. But
quantitatively the resonance masses are different in different
models. The resonance mass in QDCSM1 is about 60 MeV lower than that
in ChQM, and the QDCSM2 always has the lowest mass. For QDCCM, the
resonance mass is 2443 MeV in QDCCM0, 2298 MeV in QDCCM1 and 2156
MeV in QDCCM2. This resonance ($IJ = 03$ $\Delta\Delta$) is a
promising candidate for the observed isoscalar ABC structure seen
more clearly in the $pn \rightarrow d\pi\pi$ production cross
section at 2.36 Gev in the recent report by the CELSIUS-WASA
Collaboration \cite{ABC}.

(3) $I=1, J=0$: The $^{1}S_{0}^{\Delta\Delta}$ state is
qualitatively similar to the $^{3}S_{1}^{\Delta\Delta}$ state, since
they are just different spin-isospin states of the same quark system
with the same relative orbital angular momenta. The calculated phase
shifts, shown in the down left corner of Fig. 9, show that the
resonance survives only in QDCSM2 after the channel coupling. The
situation is almost the same as the $^{3}S_{1}^{NN}$ state.

(4) $I=1, J=2$: The phase shifts of $NN$ scattering are shown in the
down right corner of Fig. 9. From the curves, we find that a
resonance appears in QDCSM2, QDCCM1 and QDCCM2. The resonance masses
are: $2168$ MeV in QDCSM2, 2144 MeV in QDCCM1 and 2130 MeV in
QDCCM2. For ChQM and QDCSM1, only a prominent cusp appears at the
$N\Delta$ threshold. Nevertheless, the state might correspond to the
resonance looping in the Argand diagram of the $^1D_2$ $pp$-partial
wave \cite{nd2+}.

For odd-parity $NN$ states, resonance poles are found for the
isovector odd-parity $NN$ partial waves $^{3}P_{2}$, $^{3}F_{2}$ and
$^{3}F_{3}$~\cite{FA91}. These empirical resonance-like solutions
reproduce the empirical Argand loopings of the partial wave
solutions, but many studies in the past~\cite{PW1} have not resolved
the question of whether these Argand loopings represent real
dibaryon resonances. In our quark model calculation, we have not
found any resonance attributable to an $N\Delta$ or $\Delta\Delta$
bound state in the odd-parity $NN$ states.

\section{Summary}

By including the hidden color channels and varying the strength of
the color confinement potential between color-singlet channels and
hidden color channels and/or hidden color channels and hidden color
channels, a phenomenological quark model for baryon-baryon
interaction is constructed. The model achieves a good description of
$S$-,$P$-,$D$-,$F$-,$G$-,$H$-,$I$-partial wave phase shifts of $NN$
scattering as good as other quark models. It also reproduces the
binding energy and D-wave component of deuteron but a little too
small root mean square radius. Applying the model to dibaryon
search, similar results with QDCSM and ChQM are obtained. The
results show that the hidden color channels are important for the
$NN$ intermediate range attraction. The lattice QCD calculations
obtained the string like multi-body confinement interaction in the
multi-quark system~\cite{lat}. It is equivalent to the two body
confinement Eq.(4) with $k=1$ for a color singlet nucleon with three
quarks. Oka extended the string-flip model to six-quark system and
obtained a reasonable description of $NN$ interaction~\cite{Oka}
which might be viewed as a modeling of the lattice QCD string like
multi-body confinement. QDCCM fits the $NN$ scattering data better
and we suspect it might be another modeling of the lattice QCD
multi-body confinement.

Certainly one would expect to directly use the string like
multi-body interaction obtained in lattice QCD to calculate the $NN$
interaction. However, it is not only because of the huge numerical
task but also because there is no any information about the
transition interaction between different string structure which
hindered this approach.

Nuclear force is an old topic, it has been studied over 70 years and
a large amount of experimental data has been accumulated. Although
there are several approaches which can give almost perfect
description of the experimental data, the mechanism for the
intermediate-range attraction is still an open question. Lattice QCD
achieved a qualitative description of the $NN$ interaction already
and it will finally achieves a quantitative description. But based
on present lattice QCD technique it can not reveal the physical
mechanism, for example to distinguish the phenomenological $\sigma$
meson exchange and the nucleon distortion similar to molecular
covalent bond mechanism for the intermediate range attraction. One
has to develop the non-perturbative continuous QCD field theory
method as well as non-perturbative QCD model to explore the $NN$
interaction.

\end{document}